\documentclass[preprint,amsmath,amssymb,aps,fleqn]{revtex4}
\usepackage{graphicx}
\usepackage{dcolumn}
\usepackage{bm}

\begin{document}

\title{ Calculation of absolute free energy of binding for
  theophylline and its analogs to RNA aptamer using nonequilibrium
  work values }

\author{
Yoshiaki Tanida
}
\email{tanida@labs.fujitsu.com}
\author{
Masakatsu Ito
}
\author{
Hideaki Fujitani
}
\affiliation{
Fujitsu Laboratories Ltd., 10-1 Morinosato-Wakamiya, Atsugi, Kanagawa, Japan
}
\date{\today}

%%%%%%%%%%%%
% Abstract %
%%%%%%%%%%%%
\begin{abstract}
  The massively parallel computation of absolute binding free energy
  with a well-equilibrated system (MP-CAFEE) has been developed [H.
  Fujitani, Y. Tanida, M. Ito, G. Jayachandran, C. D. Snow, M. R.
  Shirts, E. J. Sorin, and V. S. Pande, J. Chem. Phys. ${\bf 123}$,
  084108 (2005)]. As an application, we perform the binding affinity
  calculations of six theophylline-related ligands with RNA aptamer.
  Basically, our method is applicable when using many compute nodes to
  accelerate simulations, thus a parallel computing system is also
  developed. To further reduce the computational cost, the adequate
  non-uniform intervals of coupling constant $\lambda$, connecting two
  equilibrium states, namely bound and unbound, are determined. The
  absolute binding energies $\Delta G$ thus obtained have effective
  linear relation between the computed and experimental values. If the
  results of two other different methods are compared, thermodynamic
  integration (TI) and molecular mechanics Poisson-Boltzmann surface
  area (MM-PBSA) by the paper of Gouda $et~al$ [H. Gouda, I. D. Kuntz,
  D. A. Case, and P. A. Kollman, Biopolymers ${\bf 68}$, 16 (2003)],
  the predictive accuracy of the relative values $\Delta\Delta G$ is
  almost comparable to that of TI: the correlation coefficients (R)
  obtained are 0.99 (this work), 0.97 (TI), and 0.78 (MM-PBSA). On
  absolute binding energies meanwhile, a constant energy shift of
  $\sim$ -7 kcal/mol against the experimental values is evident. To
  solve this problem, several presumable reasons are investigated.
\end{abstract}

\pacs{Valid PACS appear here}% PACS, the Physics and Astronomy
                             % Classification Scheme.

\maketitle

%%%%%%%%
% Main %
%%%%%%%%
\section{Introduction}
To accurately compute the free energy difference between two thermal
equilibrium states ($\Delta G$) is of interest in computational
science and of importance in terms of drug
discovery~\cite{Jorgensen2004}. It can help us obtain a quantitative
understanding of molecular complexes from atomic-scale, and also
provide valuable information for use in structural refinement for drug
design. Generally speaking, to discuss chemical reaction, we have to
determine the free energy change within so-called chemical accuracy
($\sim$ 1 kcal/mol), which is very challenging. Popularly, several
free energy calculations have been carried out within the framework of
the thermal equilibrium approach.  For instance, free energy
perturbation (FEP)~\cite{Zwanzig1954} and thermodynamic integration
(TI)~\cite{Kirkwood1935} have been widely used to calculate the free
energy change associated with the transformation of one ligand into
another via thermodynamic cycles~\cite{Tembe1984,Wong1986}.
Certainly, the recent improvement in these methods and the increase in
computer resources have enabled us to calculate the absolute binding
energy concerned with very small organic compounds~\cite{Hermans1997,
  Gilson1997, Boresch2003, Hamelberg2004}. However, equilibrium
approaches involving most rigorous technique, known as the double
decoupling method, require long time simulation to maintain the
reversibility about the work associated with the decoupling process.
As a more approximate method, the molecular mechanics
Poisson-Boltzmann surface area (MM-PBSA) analysis~\cite{Massoval1999}
has been a popular method to obtain absolute binding free energies
within a reasonable time. In this approach, explicit waters and mobile
counter ions are treated by the continuum model after obtaining the
molecular dynamics (MD) trajectory. Since the normal mode analysis is
time-consuming, the contribution of solute entropy is often
approximated based on the average of a few snapshots. On the other
hand, in the nonequilibrium statistical approach, Jarzynski recently
derived a nonequilibrium work theorem that is valid in far from
equilibrium
regime~\cite{Jar1997,Jarzynski1997-2,Crooks1999,Crooks2000,Liphardt2002,Ritort2003,Collin2005,Bustamante2005},
although accurate numerical estimates of the exponential average,
based on a finite number of sampled works, are relatively
difficult~\cite{Hendrix2001}. Shirts $et~al$. proposed the acceptance
ratio (AR)
method~\cite{Bennett1976,Cummings1998,Shirts2003,Shirts20051} to
estimate the free energy difference and also calculated the binding
free energies for eight FKBP12-ligand complexes~\cite{Shirts20052}
using the ``folding@home'' system.  Furthermore, our previous
work~\cite{Fujitani2005} suggested the efficiency of the use of
general AMBER force field (GAFF)~\cite{Wang2004} for ligands to obtain
the predictive binding estimates, and simultaneously indicated the
importance of starting structures for simulations. In our method the
use of many independent compute nodes or a grid computing system is
effective to accelerate simulations, whereupon we also developed a
massively parallel computing system, ``BioServer'', consisting of 1920
microprocessor elements, to efficiently perform the binding free
energy calculations. Hereafter, our method is referred to as the
massively parallel computation of absolute binding free energy with a
well-equilibrated system (MP-CAFEE).

The principal aim of this study is to explore the practical
feasibility of our methodology, based on a nonequilibrium approach. We
have performed the calculation of the absolute binding free energy for
theophylline and its analogs to the RNA aptamer. On the same system,
Gouda $et~al$. reported the results of free energy calculations using
two different methods, TI and MM-PBSA~\cite{Gouda2003}. Therefore,
this system is a good platform to compare the ability between two
different methods, namely the equilibrium and nonequilibrium approach.
The paper is organized as follows. We initially describe the
theoretical background and simulation protocol of binding free energy
estimation used through this work, before subsequently explaining the
choice of coupling constant $\lambda$ intervals connecting two
equilibrium states in Sec.  II. The results for all RNA-ligands are
presented and discussed in Sec. III.  After pointing out the key
features of this system, we conclude in Sec.  IV.
 
\section{Computational details}
In general, the methods used to calculate the free energy difference
in molecular simulations can be classified as either equilibrium or
nonequilibrium approaches. Several approaches based on thermal
equilibrium, e.g. FEP, TI and the potential of mean force using
weighted histogram analysis method with umbrella
sampling~\cite{Kumar1992}, have been attempted to compute the free
energy difference between two equilibrium states.  However, the
equilibrium approaches encounter the difficulty of removing the
nonequilibrium contributions, known as the hysteresis
problem~\cite{Straatsma1986,Perlman1989-1,Perlman1989-2}. In order to
overcome this difficulty, a long time simulation must be performed, to
retain the reversibility of quasistatic process associating with the
thermal equilibrium. Recently, Jarzynski's equality exactly relates
the free energy difference of two equilibrium states $i$ and $j$ to
the statistics of works via a nonequilibrium, irreversible process
that connects them as $\displaystyle{\rm exp}(-\Delta G/k_{\rm B}T)=
\langle {\rm exp}(-W/k_{\rm B}T)\rangle$, where $W$ is the work change
of a system from $i$ to $j$ under isothermal isobaric (NPT)
conditions~\cite{Evans0611541}. A remarkable feature of Jarzynski's
equality is the fact that the switching time between two states is
arbitrary. Subsequently, when the process occurs instantaneously, the
work can be defined by the potential energy difference as $W\equiv
U(\lambda_{\rm j}, {\bf x})-U(\lambda_{\rm i}, {\bf x})$, where $U$
and $\bf x$ are the potential function and configurational
coordinates, respectively. A coupling constant $\lambda$ is imposed on
the path connecting two states. The fluctuation theorem (FT) related
with $P_{{\rm i}\to{\rm j}}(W)$ and $P_{{\rm j}\to{\rm i}}(W)$ of the
work probability distributions along the non-equilibrium forward
($i\rightarrow j$ direction) and reverse ($j\rightarrow i$ direction)
processes is derived by Crooks~\cite{Crooks2000} as $P_{{\rm i}\to{\rm
    j}}(+W)/P_{{\rm j}\to{\rm i}}(-W)= {\rm exp}[(W-\Delta G)/k_{\rm
  B}T]$. The ratio between $P_{{\rm i}\to{\rm j}}(+W)$ and $P_{{\rm
    j}\to{\rm i}}(-W)$ depends only the value of the free energy
difference $\Delta G$.  Shirts $et~al$. pointed out that the maximum
likelihood $\Delta G$ is given by AR analysis, namely, the control
parameter in AR analysis must correspond to the free energy
difference~\cite{Shirts2003}. Since two irreversible work
distributions in forward and reverse directions cross at $W$= $\Delta
G$, both distributions must have a large overlap to obtain the free
energy estimate with high accuracy. Subsequently, after applying the
multi-staging method, the intermediate stages are inserted onto the
path between two states $i$ and $j$, with the total free energy change
obtained as the sum of the differences. This type of usage, involving
the insertion of many intermediate states on the path connecting two
states, seems to be effective. However, since economical issues also
play a role because many $\lambda$ require significant computing
resources, we have now addressed the issue concerned with determining
the adequate choice of each $\lambda$ interval. First, potential
energy is parameterized by $\lambda$, followed by the introduction of
the $\lambda$ dependency to the non-bonded interactions. In order to
avoid instability near the end-points of the vanishing atoms through
calculation, the non-bonded interaction energy $U(\lambda^{\rm
  C},\lambda^{\rm LJ})$ between the ligand and others, including the
so-called soft-core potentials, is used~\cite{Shirts2005-2}.  We also
use two kind of coupling constant $\lambda$ for representing two
different potential energies: the fully bound state ($\lambda^{\rm C}=
\lambda^{\rm LJ}= 0$) and the unbound state ($\lambda^{\rm C}=
\lambda^{\rm LJ}= 1$).  Hereafter, the free energy component resulting
from turning off the electrostatic potential ($\lambda^{\rm C}= 0\to
1, \lambda^{\rm LJ}= 0$) is referred as the electrostatic
contribution. We also denote the free energy component for the process
of ($\lambda^{\rm C}= 1, \lambda^{\rm LJ}= 0\to 1$) as the van der
Waals contribution. Next, we investigate the convergence properties of
the free energy when the number of equal $\lambda$ spacings ($\equiv
N$) increases. A trajectory at each intermediate $\lambda_{i}$ is
obtained via the standard MD simulation. As already noted, the work
associated with the forward path is obtained to compute the potential
energy difference as $W_{\rm F}(i, i+1)= U(\lambda_{i+1}, {\bf
  x}_{i})-U(\lambda_{i}, {\bf x}_{i})$. Using a trajectory at
$\lambda_{i+1}$, the work on the reverse path is also obtained as
$W_{\rm R}(i, i+1)= U(\lambda_{i+1}, {\bf x}_{i+1})-U(\lambda_{i},
{\bf x}_{i+1})$, while the free energy $\Delta G_{i, i+1}$ can be
estimated using both work distributions of $W_{\rm F}(i, i+1)$ and
$W_{\rm R}(i, i+1)$ by the AR algorithm, thus the free energy
difference is $\Delta G= \sum^{N-1}_{i= 0}\Delta G_{i, i+1}$. In
general, $\Delta G$ is being converged as the number of $N$ increases.
$N_{0}$ is defined as a large enough number to obtain the
well-converged value of $\Delta G$ ($\equiv \Delta G_{0}$). We also
define $\Delta G (\lambda_{n})= \sum^{n-1}_{i= 0}\Delta G_{i, i+1}$.
Finally, to reproduce the feature of $\Delta G_{0}(\lambda_{n})$, the
adequate mesh points of $\lambda$ are determined to be satisfied with
the following requirements: 1) the small difference between $\Delta G$
and $\Delta G_{0}$; 2) the small root mean square (rms) error between
$\Delta G(\lambda_{n})$ and $\Delta G_{0}(\lambda_{n})$. The rms error
used here is defined as $ \left[(1/N)\sum^{N-1}_{i= 0}\mid\Delta
  G(\lambda_{i})-\Delta G_{0}(\lambda_{i})\mid^{2}\right]^{1/2}$.

\begin{figure*}
\includegraphics[width=10cm]{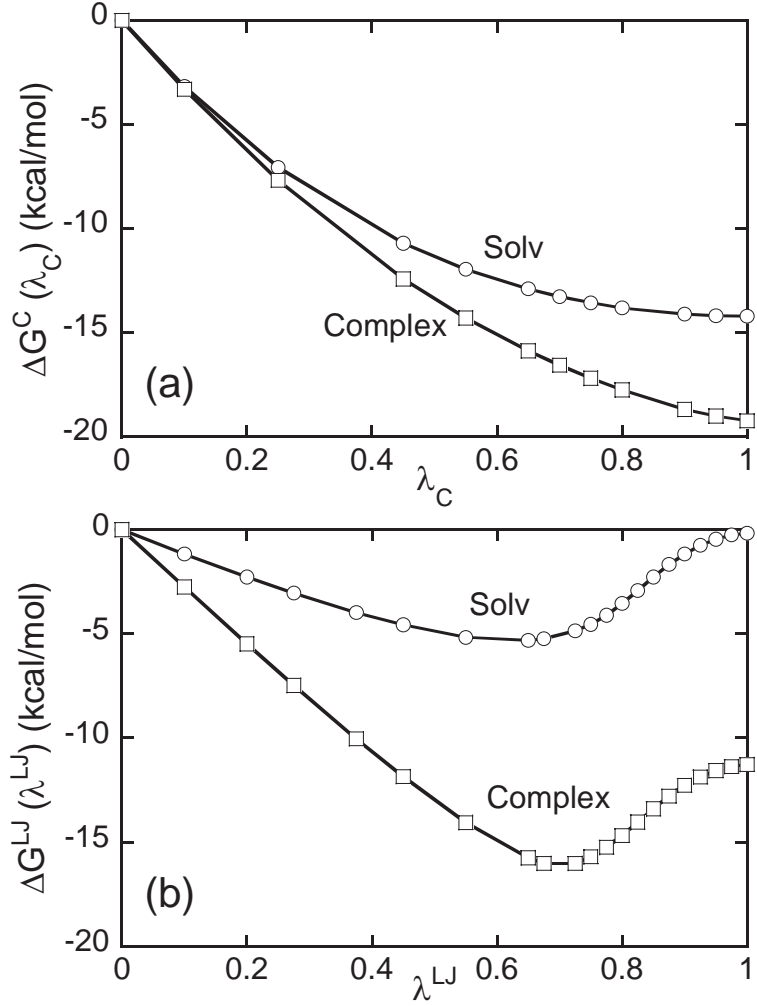}
\caption{\label{fig:Lambda_dependence} Free energy convergence as a
  function of the number of $\lambda$ spacing for the RNA-Theophylline
  system; (a) $\Delta G^{\rm C}$ vs. $\lambda^{\rm C}$. Open symbols
  are our reduced 12 $\lambda^{\rm C}$ points of (0, 0.1, 0.25, 0.45,
  0.55, 0.65, 0.7, 0.75, 0.8, 0.9, 0.95, 1). The solid lines are
  generated from 20 uniform mesh points of $\lambda^{\rm C}$. (b)
  $\Delta G^{\rm LJ}$ vs.  $\lambda^{\rm LJ}$. Open symbols are our
  reduced 21 $\lambda^{\rm LJ}$ points of (0, 0.1, 0.2, 0.275, 0.375,
  0.45, 0.55, 0.65, 0.675, 0.725, 0.75, 0.775, 0.8, 0.825, 0.85,
  0.875, 0.9, 0.925, 0.95, 0.975, 1), while the solid lines are
  generated from 40 uniform mesh points of $\lambda^{\rm LJ}$.}
\end{figure*}

Now, we have applied this manner to an RNA-theophylline system. Using
the first 100 ps MD simulation $N_{0}$ were determined as 20 for
electrostatic contribution and 40 for van der Waals contribution
respectively. As can be seen from Table~\ref{tab:conv_lambda}, we have
determined the non-uniform mesh points of $\lambda$ with $\Delta G$
difference ($\equiv \delta G$) $\leq$ 0.05 kcal/mol and rms error of
$\Delta G(\lambda)$ $\leq$ 0.1 kcal/mol, leading to a significant
reduction in the number of necessary $\lambda$ points. The $\lambda$
points obtained were the following: 12 values of $\lambda^{\rm C}$
points (0, 0.1, 0.25, 0.45, 0.55, 0.65, 0.7, 0.75, 0.8, 0.9, 0.95, 1)
and 21 values of $\lambda^{\rm {LJ}}$ points (0, 0.1, 0.2, 0.275,
0.375, 0.45, 0.55, 0.65, 0.675, 0.725, 0.75, 0.775, 0.8, 0.825, 0.85,
0.875, 0.9, 0.925, 0.95, 0.975, 1). Computed $\Delta G(\lambda_{i})$
as a function of $\lambda_{i}$ were also shown in
Figs.~\ref{fig:Lambda_dependence}(a) and (b). It is clear that the feature of
the well-converged curves is reproduced well by a small number of
non-uniform mesh points.

To estimate the free energy difference, the path from the bound state
($\lambda$= 0) to the unbound state ($\lambda$= 1) is chosen as
follows; the ligand electrostatic interactions with the environment
are turned off, followed by its van der Waals interactions with the
environment, whereupon we obtain the solvation energy of ligand as
$\Delta G_{\rm Solv}$ ($\equiv \Delta G_{\rm Solv}^{\rm C}+\Delta
G_{\rm Solv}^{\rm LJ}$). The free energy of the annihilation of the
ligand from the ligand-receptor complex (complexation free energy) is
also calculated as $\Delta G_{\rm Complex}$ ($\equiv \Delta G_{\rm
  Complex}^{\rm C}+\Delta G_{\rm Complex}^{\rm LJ}$). We schematically
note as follows; $L_{(\rm sol)}\rightarrow L_{(\rm gas)}$ on the
calculation of the solvation free energy of the ligand (L) and
$RL_{(\rm sol)}\rightarrow R_{(\rm sol)}+L_{(\rm gas)}$ on the
calculation of the complexation free energy of the ligand (L) to the
receptor (R).  Consequently, we obtain the absolute binding free
energy of the receptor-ligand complex as $\displaystyle\Delta G \equiv
\Delta G_{\rm Complex}-\Delta G_{\rm Solv}=\Delta G_{\rm Complex}^{\rm
  C}+\Delta G_{\rm Complex}^{\rm LJ}-\Delta G_{\rm Solv}^{\rm
  C}-\Delta G_{\rm Solv}^{\rm LJ}$ using the relation of $RL_{(\rm
  sol)}\rightarrow R_{(\rm sol)}+L_{(\rm sol)}$.

\begin{figure*}
\includegraphics[width=14cm]{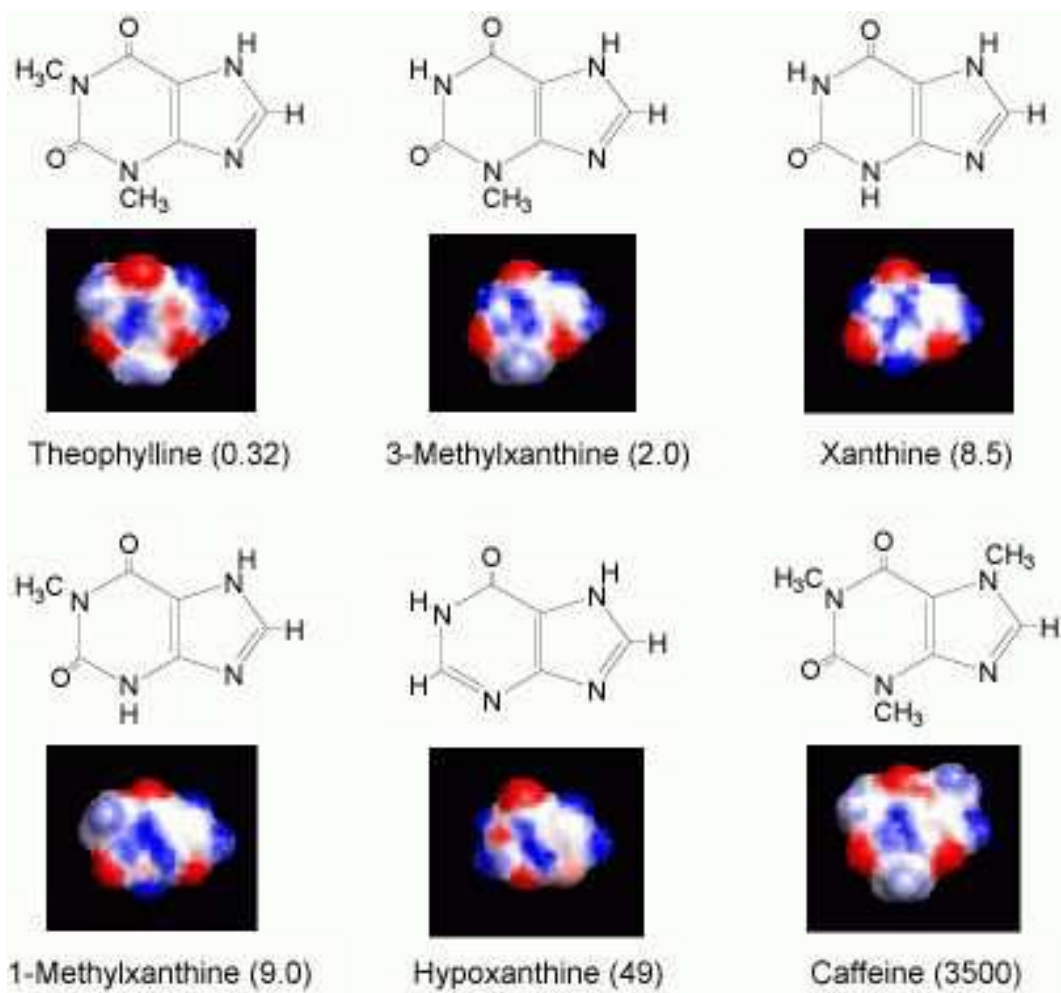}
\caption{\label{fig:6Ligand_APBS} Chemical structures of ligands
  investigated. The dissociation constant $K_{\rm d}$ (in $\mu \rm M$)
  is included in each parenthesis. The electrostatic potential
  surfaces of molecules are also shown with red representing the
  negative potential and blue representing the positive potential
  respectively (-4 $k_{\rm B}T/e$ to 4 $k_{\rm B}T/e$).  }
\end{figure*}

In our study the three-dimensional structure of the RNA-theophylline
complex proposed by Clore $et~al$.~\cite{Clore2003} (PDB code 1O15)
was used as a modeling. We assumed that the binding sites for six
ligands fit into the structural framework of the original
RNA-theophylline structure because the structures of other complexes
were not experimentally determined. All MD simulations were performed
using a modified version of the GROMACS package
(v3.1.4)~\cite{Lindahl2001} with single precision to speed up the
execution. The Amber force field parameters (ff99) were used to
describe the RNA aptamer, including 1-4 interactions between hydrogen
atoms, while the force field parameters for Mg$^{2+}$
ions~\cite{Aqvist1990} were also used from the Amber package. The
atomic structure of the theophylline molecule was optimized in a
vacuum using Gaussian 98 (Gaussian, Inc.)  with the HF/6-31G$^{*}$
single-point energy calculation, while the atomic charges were
calculated using the restrained electrostatic potential (RESP)
method~\cite{Bayly1993} and the ANTECHAMBER program was used to assign
the GAFF for each atom. The chemical structures of the ligands
investigated here are shown in Fig.~\ref{fig:6Ligand_APBS}, while the
electrostatic potential surface of these molecules, computed using the
Adaptive Poisson-Boltzmann Solver (APBS) package~\cite{Baker2001}
after using the RESP charge fitting are also shown. It is clear that
hypoxanthine has a slightly different charge distribution around the
lower left-hand side of the molecule as compared to other molecules.
The TIP3P model for water molecules was used to describe the
solvent~\cite{Jorgensen1983}.

Through this study we obeyed the following parameters for all MD
simulations. Integration of the equation of motion was performed using
the leap-frog algorithm and all bonds were constrained by
LINCS~\cite{Hess1997} with order 8. We used the time step of 2.0 fs
while the non-bonded pair list of 1 nm was updated every 10 steps. All
simulations were carried out at $T$= 298 K using the
Nose-Hoover~\cite{Nose1984,Hoover1985} temperature control and at 1
atm using the Berendsen~\cite{Berendsen1984} pressure control
respectively, with a time constant of 0.5 ps and a compressibility of
4.5$\times$10$^{-5}$ bar$^{-1}$.  The particle mesh Ewald
(PME)~\cite{Essmann1995} summation was used to evaluate the
electrostatic interactions with a grid spacing of approximately 0.12
nm, a cubic spline order of 4 and relative tolerance between long and
short range energy of $10^{-8}$.  The L-J interaction energy was
calculated by a switched cutoff between 0.8 nm and 0.9 nm. We also
included long range correction to the energy and pressure. We used the
truncated octahedron box as a unit cell with periodic boundary
conditions, while for the solvated ligand system, we adopted the unit
cell size, maintained no closer than 0.85 nm from each face of the
box, and introduced about 238 to 282 TIP3P water molecules into the
unit cell. RNA-ligand complexes were prepared in which the minimum
distance was 0.75 nm between the nearest unit cell wall on either
side, while the number of TIP3P water molecules used were either 7192
or 7193. We introduced 3 Mg$^{2+}$ ions into the complex according to
the paper of Gouda
$et~alt$~\cite{Jenison1994,Zimmermann2000,Gouda2003} and also added 26
Na$^{+}$ counter-ions to the complex to neutralize the system.

One of the difficulties for computing the binding free energy is the
preparation of starting structure sampled from the canonical ensemble
at the desired temperature $T$~\cite{Fujitani2005}. The following
procedure was used. We performed energy minimization using the
conjugate gradient method, followed by the MD simulation of 20 ps
relaxation with the solute positions restrained. In order to attain
the thermal equilibrium state ($\lambda^{\rm C}$=$\lambda^{\rm LJ}$=
0), 10/15 ns MD calculations were typically used for the solvated
ligand and the RNA-ligand complexes. Through this process, used to
conclude the system to be equilibrated, the total potential energy of
the system and the local electrostatic and van der Waals potential
energies between RNA and ligand were carefully monitored. Via the
multi-stage method we adopted an atomic structure at $\lambda$= 0
after completing the preconditioning MD simulation as a starting
structure at each intermediate $\lambda$ state. To decrease the
uncertainty from statistical error, we sampled the MD simulation with
15 and 10 kinds of different initial velocities for the solvated
ligand system and RNA-ligand complex, respectively. Nonequilibrium
works every 50 steps (0.1 ps) were used to compute the free energies,
which were obtained as the arithmetic average of all estimated values.
Through this work we simultaneously performed 330 (33 independent
non-uniform $\lambda$ points $\times$ 10 samples) independent MD
simulations to generate the energetically possible samples for
evaluating the binding free energy on the Fujitsu BioServer system,
consisting of 1920 FR-V microprocessors.

\section{Results and discussion}
It was found that a Mg$^{2+}$ ion shielded by water molecules and
located at the center of the U-turn formed by
C22-U24~\cite{Zimmermann2000} remained unchanged during all the
preconditioning MD simulations. Conversely however, other Mg$^{2+}$
ions immediately left their initial positions, and diffused around an
RNA aptamer.

\subsection{Theophylline}
A 15 ns preconditioning MD simulation was used to reach a thermal
stable structure in the RNA-theophylline complex. We now consider the
energetically favorable structure from the perspective of both
electrostatic interaction and that of van der Waals. The obtained
structure is consistent as expected, namely, the negative surface
region of an RNA strongly interacts with the positive region on a
complementary molecule surface. Besides, there is a match of
theophylline and residue of A28 on the substrate. This result can be
interpreted based on the view that the van der Waals interaction
simultaneously stabilizes the theophylline molecule onto the binding
pocket. Furthermore, theophylline is bound with an upper as well as
lower layer.

\begin{figure}
\includegraphics[width=12cm]{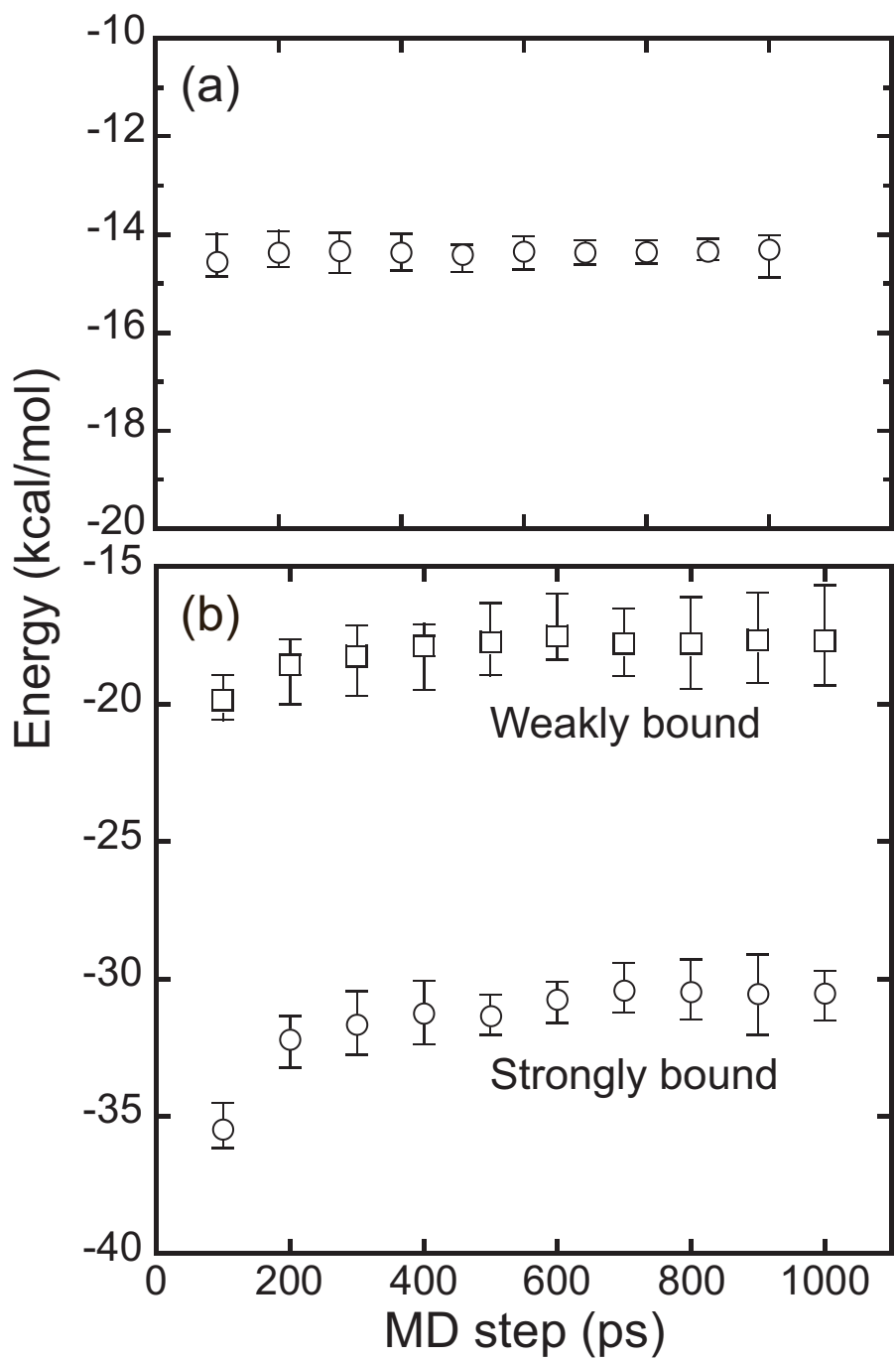}
\caption{\label{fig:FE-TEP-RNA} Free energy of theophylline-RNA
  aptamer as a function of MD step: (a) $\Delta G_{\rm Solv}$ for
  solvated theophylline, (b) $\Delta G_{\rm Complex}$ for the
  RNA-theophylline complex. Open circle denotes a strongly binding
  structure, while the open square denotes a weakly binding structure
  (see, in IIID). Error bars in both graphs are drawn from the maximum
  value to minimum value.}
\end{figure}

Let us now move on to the binding free energy calculation to examine
the availability of our method. The open circles in
Fig.~\ref{fig:FE-TEP-RNA} show the features of the convergence of (a)
the solvation free energy of theophylline and of (b) the complexation
free energy of theophylline to an RNA as a function of the MD step.
Each open circle was obtained by analyzing the generated works during
each 100 ps MD trajectory. Rough convergence appears visible after 200
ps and 700 ps for (a) the solvated ligand and (b) the solvated
RNA-ligand complex, respectively. Subsequently, we used work values
from 200 ps to 1 ns (8000 samples) to estimate the solvation energy
$\Delta G_{\rm Solv}$ and those from 700 ps to 1 ns (3000 samples) to
estimate the complexation energy $\Delta G_{\rm Complex}$, and
obtained -14.4 kcal/mol for $\Delta G_{\rm Solv}$ and -30.5 kcal/mol
for $\Delta G_{\rm Complex}$. Consequently, we also obtained the
binding free energy $\Delta G$ of -16.1 kcal/mol for the
RNA-theophylline system.  We have summarized the free energy
components in Table~\ref{tab:components} and it can be found that the
favorable structure of this complex is mainly driven by the van der
Waals interaction.

We further comment on the computational cost required to obtain the
converged $\Delta G$. Since the ratio of the simulation time between
the solvation and complexation is relatively small, we only consider
the simulation time for obtaining $\Delta G_{\rm Complex}$. In this
system, a total of 330 ns MD simulations for obtaining $\Delta G_{\rm
  Complex}$ were used. However, our approach is applicable for the use
of many compute nodes to accelerate calculations, meaning the use of a
parallel computing system, including a grid environment, is the key to
diminishing the computation turnaround time.

\subsection{Caffeine}
\begin{figure}
\includegraphics[width=14cm]{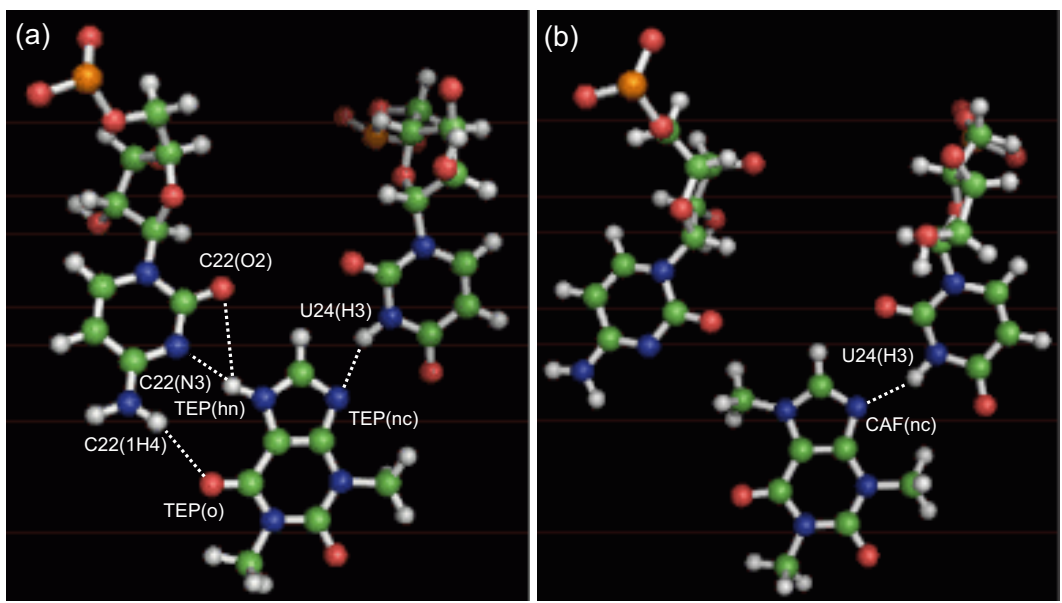}
\caption{\label{fig:local_struct} (a) Theophylline molecule and its
  surroundings in the RNA-theophylline complex. Only two residues
  related to hydrogen bonds are drawn and candidates for hydrogen
  bonds are depicted as dashed lines. (b) Caffeine molecule and its
  surroundings in the RNA-caffeine complex, while hydrogen bonds with
  residue in the same plane are also drawn in the form of a dashed
  line.}
\end{figure}

Caffeine differs from theophylline by a single methyl group as shown
in Fig.~\ref{fig:6Ligand_APBS}. Each component, once decomposed from
estimated binding energies, was tabulated in
Table~\ref{tab:components}. It is evident that the significant
difference of 5.1 kcal/mol between the theophylline and caffeine
bindings is mainly attributable to the electrostatic contribution (4.6
kcal/mol). We now address and attempt to explore the reason why the
affinity between RNA and caffeine is drastically reduced in comparison
with that between RNA and theophylline. Due to the typical snapshots,
both ligands preferably bind within the pocket of an RNA, as
illustrated in Figs.~\ref{fig:local_struct}(a) and (b). Hereafter, we
discuss the results during 1ns MD simulations after reaching thermal
equilibrium states for both systems. The obtained averaged distances
of TEP(O)-C22(1H4), TEP(hn)-C22(N3), TEP(hn)-C22(O2) and
TEP(nc)-U24(H3) are 0.22 nm, 0.19 nm, 0.26 nm and 0.21 nm,
respectively, while the averaged distance of CAF(nc)-U24(H3) of 0.21
nm was also obtained for the RNA-caffeine complex. These values
indicated the creation of hydrogen bonds as the bond lengths were less
than approximately 0.25 nm. This difference in the number of hydrogen
bonds between both RNA-caffeine and RNA-theophylline complexes may
result in a significant difference in the binding energy, so the
number of hydrogen bonds for both systems were enumerated. At this
point, we determined the hydrogen bond based on the cutoff angle of
the donor-hydrogen-acceptor and the distance of the hydrogen-acceptor;
values of 60$^{\circ}$ for cutoff angle and 0.25 nm for cutoff
distance were used. We obtained hydrogen bonds of 4.3 ($\pm$1.0) for
the solvated caffeine and the numerical value in parentheses indicated
the standard deviation. On the other hand, the solvated theophylline
has 5.4 ($\pm$1.1) hydrogen bonds. These results indicated a decrease
in the hydrogen bond in solvated caffeine as compared to the solvated
theophylline. The number of hydrogen bonds for the solvated
RNA-caffeine complex were 2.8 ($\pm$0.8) and 0.9 ($\pm$0.2) for
caffeine-water and RNA-caffeine pairs. We also counted the number of
hydrogen bonds for the solvated RNA-theophylline complex: 2.3
($\pm$0.7) for theophylline-water and 3.1 ($\pm$0.6) for
RNA-theophylline. This value ($\sim$ 3) is easily understood as the
hydrogen atom TEP(hn) are shared to be bound by both a nitrogen atom
(C22(N3)) and an oxygen atom (C22(O2)) in the solvated
RNA-theophylline complex as shown in Fig.~\ref{fig:local_struct}(a),
while these hydrogen bonds vanish in the solvated RNA-caffeine complex
in Fig.~\ref{fig:local_struct}(b). The hydrogen bond contributing to
binding energy for the RNA-theophylline system is $\sim$ 0, while the
decrease in the number of hydrogen bonds for the RNA-caffeine system
is $\sim$ 0.6. The contribution energies to binding free energies are
roughly $\sim$ 0 kcal/mol for the RNA-theophylline system and $\sim$
3.6 kcal/mol for the RNA-caffeine system, based on a presumed hydrogen
bond energy of $\sim$ 6 kcal/mol. Therefore, the decrease in the
hydrogen bonds represents the main reason for the significant
difference in affinity between the RNA-theophylline and RNA-caffeine
systems.

\subsection{Other molecules}
To obtain a thermal equilibrium binding structure, we similarly set
the initial locations of other molecules in close proximity to the
binding site of theophylline for preconditioning MD simulations
because the electrostatic potential surface of each molecule is almost
the same as theophylline. The obtained stable atomic configurations of
other molecules bound with RNA were virtually unchanged.  The
association of RNA with all six ligands is driven primarily by
attractive van der Waals interactions $\Delta G^{\rm LJ}$ as tabulated
in Table~\ref{tab:components}. The results of the absolute and
relative binding free energies we obtained for six ligands are given
in Table~\ref{tab:FE_6ligands}, which also tabulates the results of
other methods, TI and MM-PBSA~\cite{Gouda2003} and includes three
important characteristics: 1) There is an effective linear relation
(slope= $1$) between the computed absolute binding free energies and
experimental ones. We obtained the following relation between the
computed relative energies $\Delta\Delta G_{\rm calc}$ and
experimental ones $\Delta\Delta G_{\rm expt}$ for three different
methods (the results of TI and MM-PBSA were from
Ref.~\cite{Gouda2003}): $\Delta\Delta G_{\rm calc}= \Delta\Delta
G_{\rm expt}-0.1, R= 0.99$ (This work), $\Delta\Delta G_{\rm calc}=
\Delta\Delta G_{\rm expt}+0.19, R= 0.97$ (TI), $\Delta\Delta G_{\rm
  calc}= \Delta\Delta G_{\rm expt}+1.68, R= 0.78$ (MM-PBSA), where $R$
is the correlation coefficient. Consequently, the efficiencies of both
our method and the TI method for estimating the relative binding free
energies are virtually comparable. Here, we should point out that our
method directly estimate the absolute binding energies, whereas the TI
method only compute the relative binding energies. We also point out
that the absolute binding free energy for the RNA-hypoxanthine system
is partly shifted about 1.5 kcal/mol from the linear relation, which
means this result might suggest consideration be taken of another
binding site for hypoxanthine binding. 2) There is a constant energy
shift of $\sim$ -7 kcal/mol from the experimental values $\Delta
G_{\rm expt}$ as can be seen in Fig.~\ref{fig:DG_plot_6molecules}.
Detailed discussions for the latter are given below.

\begin{figure}
\includegraphics[width=14cm]{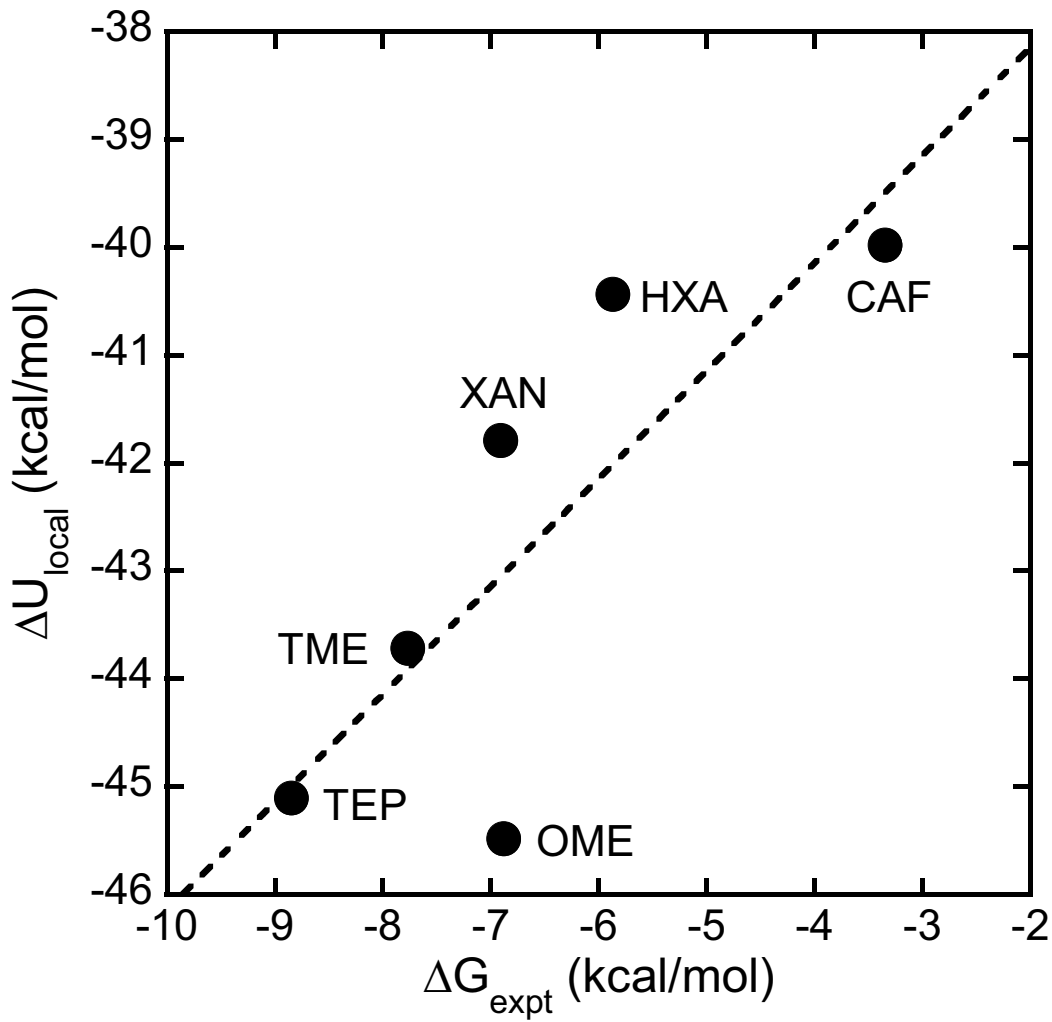}
\caption{\label{fig:Potential-DG_expt} Local potential energies
  computed $\Delta U_{\rm local}$ vs. experimental binding free
  energies $\Delta G_{\rm expt}$ for six ligands. The line of slope= 1
  is also drawn as a guide, while the abbreviations used are the same
  as those in Fig. 5.  }
\end{figure}

In general, since water molecules play an important role in free
energy calculation, we have performed calculations of $\Delta U_{\rm
  local}$ as the sum of short-range electrostatic interaction energy
and van der Waals interaction energy between RNA and ligand. The
sample averaged values during 1 ns MD simulations after reaching the
thermal equilibrium state are shown in
Fig.~\ref{fig:Potential-DG_expt}. In comparison with
Fig.~\ref{fig:DG_plot_6molecules}, our method can reproduce the trend
in both 1-methylxanthine and xanthine well, although the large shift
($\simeq$ 2 kcal/mol) from a linear relation between $\Delta U_{\rm
  local}$ and $\Delta G_{\rm expt}$ was evident.  Consequently, our
method evidently takes the effect of water molecules into
consideration.

\subsection{Constant energy shift}
The computed absolute binding free energies were approximately -7
kcal/mol smaller than those obtained by the experiment. How do we
solve the puzzle underlying the difference between them? We briefly
comment on the presumed reasons to solve this puzzle. First, it is
well known that nucleic acid, e.g. RNA or DNA, is generally flexible
in solution ~\cite{Sorin2004}. There are a number of metastable states
in free energy space and long term residence at a single valley,
resulting in anomalously slow dynamics. Namely, the ensemble average
of several capable atomic structures was observed in the experiment.
Thus, in the RNA-theophylline system we examined exploration of
another metastable structure using a simulated annealing technique as
follows: We first prepared a slightly expanded unit cell that was
sufficient to treat the RNA relaxing. Theophylline and RNA aptamer was
solvated in a truncated octahedron box with 3 Mg$^{2+}$ ions, 26
Na$^{+}$ counter-ions and 9325 TIP3P water molecules. Next, energy
minimization and position restrained MD simulation were carried out in
the same manner, as described in II. We also set a simulation
temperature of 398 K to disturb the stable complex structure during
the simulation time of 100 ps. Consequently, a (meta)stable atomic
configuration was obtained by annealing to 298 K. One of the
characteristic features was that a Mg$^{2+}$ ion leave from the center
of the U-turn of C22-U24. Binding free energy of -3.3 kcal/mol was
obtained by AR analysis with simulation of up to 1 ns MD, by
discarding the first 400 ps runs, which is significantly small
compared with the strongly binding structure described in Sec. IIIA.
The feature of convergence of the free energy estimation as a function
of the MD step for this weakly binding structure was also drawn in
Fig.~\ref{fig:FE-TEP-RNA}.  We further divide the binding free energy
into electrostatic and van der Waals components. Binding free energy
of -0.4 kcal/mol is contributed from the electrostatic part, whereas,
-3.0 kcal/mol is contributed from the van der Waals part. Second, we
examined the influence of the atomic charges of the ligand. Here,
AM1-BCC charges~\cite{Jakalian2000} of the ligand generated from
MOPAC2002 were tested. $\Delta G_{\rm Solv}$ of -17.7 kcal/mol was
obtained, while $\Delta G_{\rm Complex}$ of -33.7 kcal/mol was also
obtained up to 1 ns MD simulations. Both values increased by $\sim$ 3
kcal/mol as compared to those using RESP charges; hence we obtained
binding energy of -15.9 kcal/mol, which is very close to the value of
-16.1 kcal/mol using RESP charges. Third, the convergence of binding
energy was explored. The solvation energy $\Delta G_{\rm Solv}^{\rm
  LJ}$ was reduced by approximately 0.5 kcal/mol as the cutoff radius
was increased from 0.9 nm to 1 nm. The $\Delta G_{\rm Complex}^{\rm
  LJ}$ was also reduced by 0.2 kcal/mol as the cutoff radius increased
from 0.9 nm to 1 nm, meaning the correction is substantially only 0.3
kcal/mol. Fourth, we previously reported that $\sim$ 3 kcal/mol energy
shift from the experimental values is existed in FKBP12-ligand
system~\cite{Fujitani2005}. Then, several researchers have tried to
solve this issue by different
approaches~\cite{Jayachandran2006,Mobley2006,Wang2006}. Currently, we
believe that this energy shift can be interpreted by the difference of
the force field used. The energy shift vanishes when we use the GAFF
for representing the potential of the protein to further improve the
consistency. Therefore, we examined computing the binding energy using
GAFF representing all atoms in the RNA-theophylline system. $\Delta
G_{\rm Solv}$ of -14.2 kcal/mol was obtained, while $\Delta G_{\rm
  Complex}$ of -27.0 kcal/mol was also obtained, so we obtained a free
energy difference $\Delta G$ of -12.8 kcal/mol. This value still
represents an approximate shift from the experimental values by 4.0
kcal/mol. Finally, since an RNA has significant negative charges, we
believe the fact that the charge polarizability of an atom is
influenced by electrostatic interactions with its surrounding atoms,
becomes important~\cite{Dang1987, Caldwell1990,Mahoney2001}. For this
reason the polarizable potential model might be considered, however,
as the computational cost rises significantly, we have not yet
examined this ability. Consequently, this puzzle remains a major
challenge in the RNA-ligand system.

\section{Conclusion}
In conclusion, we have introduced the method (MP-CAFEE) as a highly
accurate means of obtaining the absolute binding free energy. This
method has been applied to investigate the properties of the
RNA-ligand system. Since our method is suitable for the use of many
compute nodes, we have also developed a massively parallel computing
system to effectively accelerate simulations. To further reduce the
computational cost, we determined the adequate non-uniform intervals
of two coupling constants and consequently obtained the following
details: First, our estimated absolute binding free energies correlate
well in terms of a linear fit to the experimental values when all
ligands are assumed to be preferable to bind the same binding pocket.
By comparing the results of two other methods, TI and
MM-PBSA\cite{Gouda2003}, the accuracy of this method is almost
comparable to that of TI in terms of relative binding free energy.
With this in mind, we believe this method to be a promising tool to
quantitatively predict the absolute binding free energy in the drug
design domain. Second, we expect a considerable difference in the
binding affinity to RNA between theophylline and caffeine to be
attributed to the difference in hydrogen bonds of the related
environment by analyzing the free energy component. Finally, we find
the constant energy of $\sim$ -7 kcal/mol shifted from the
experimental values for all ligands on absolute binding free energy.
Though considering several presumable solutions to this problem, this
area remains incompletely understood.

\begin{acknowledgments}
  The authors would like to thank G. Jayachandran, M. R. Shirts, E. J.
  Sorin, and V. S. Pande for fruitful discussions and also E. Lindahl
  for help in modifying GROMACS. This work was partly supported by the
  High-Throughput Biomolecule Analysis System Project of the NEDO (New
  Energy and Industrial Technology Development Organization, Japan).
  Three of the authors (H.F., Y.T., and M.I.) would like to thank all
  the members of the BioServer project in Fujitsu.

\end{acknowledgments}

%%%%%%%%%%%%%%%%
% Reference(s) %
%%%%%%%%%%%%%%%%

%%\bibliography{RNA_TEP.bib}

%%%%%%%%%%%%%%%%%%%%%%
% Tables and Figures %
%%%%%%%%%%%%%%%%%%%%%%

\newpage
%%%%%%%%%%%
% Table 1 %
%%%%%%%%%%%
\begin{table*}
  \caption{
    Obtained free energy differences and rms values comparing with well-converged values using condensed equal $\lambda$ spacings, 20 uniformed points for $\lambda^{\rm C}$, and 40 uniformed points for $\lambda^{\rm LJ}$ respectively. These values are deduced from the first 100 ps MD trajectory.}
\label{tab:conv_lambda}
\begin{ruledtabular}
\begin{tabular}{lccccc}
  &&\multicolumn{2}{c}{solvation}&\multicolumn{2}{c}{complex}\\
  &\multicolumn{1}{c}{$N$}&$\delta G\footnotemark[1]$&rms&$\delta G\footnotemark[1]$&rms\\ \hline
  \multicolumn{1}{c}{$\lambda^{\rm C}$}&2&-0.14&0.09&0.08&0.18\\
  &5&0.11&0.09&0.10&0.14\\
  &10&0.03&0.03&0.06&0.07\\
  &11\footnotemark[2]&0.03&0.03&-0.01&0.03\\ \hline
  \multicolumn{1}{c}{$\lambda^{\rm LJ}$}&2&-6.45&3.73&-10.20&5.89\\
  &5&-2.69&1.16&-2.42&1.19\\
  &10&-0.28&0.13&-1.48&0.60\\
  &20&0.11&0.09&0.27&0.13\\
  &20\footnotemark[2]&0.05&0.08&-0.05&0.05\\
\end{tabular}
\end{ruledtabular}
\footnotetext[1]{$\delta G\equiv \Delta G- \Delta G_{0}$, where $\Delta G_{0}$ is well-converged value.}
\footnotetext[2]{These values are obtained using an adaptive $\lambda$ spacing set (see, in II).}
\end{table*}

\newpage
%%%%%%%%%%%
% Table 2 %
%%%%%%%%%%%
\begin{table*}
  \caption{
    Electrostatic contributions and those of van der Waals, decomposed from obtained binding free energies for each RNA-ligand complex.}
\label{tab:components}
\begin{ruledtabular}
\begin{tabular}{lccccccc}
  &\multicolumn{2}{c}{$\Delta G_{\rm Solv}$}&\multicolumn{2}{c}{$\Delta G_{\rm Complex}$}&\multicolumn{2}{c}{$\Delta G$}\\
  \multicolumn{1}{c}{Ligand}&{$\Delta G_{\rm Solv}^{\rm C}$}&{$\Delta G_{\rm Solv}^{\rm LJ}$}&{$\Delta G_{\rm Complex}^{\rm C}$}&{$\Delta G_{\rm Complex}^{\rm LJ}$}&{$\Delta G^{\rm C}$}&{$\Delta G^{\rm LJ}$}\\ \hline
  Theophylline&-14.2&-0.2&-19.2&-11.3&-5.0&-11.1\\
  3-Methylxanthine&-16.8&-0.5&-22.1&-9.9&-5.3&-9.4\\
  Xanthine&-19.1&-1.0&-24.8&-10.0&-5.7&-9.0\\
  1-Methylxanthine&-16.7&-0.6&-21.3&-10.2&-4.6&-9.6\\
  Hypoxanthine&-18.2&-0.3&-22.8&-7.2&-4.6&-6.9\\
  Caffeine&-12.7&0.3&-13.1&-10.3&-0.4&-10.6\\
\end{tabular}
\end{ruledtabular}
\end{table*}

\newpage
%%%%%%%%%%%
% Table 3 %
%%%%%%%%%%%
\begin{table*}
  \caption{
    Absolute binding free energies obtained for RNA aptamer and
    theophylline and its analogs in kcal/mol.
  }
\label{tab:FE_6ligands}
\begin{ruledtabular}
\begin{tabular}{lccccccc}
  &\multicolumn{2}{c}{Experiment\footnotemark[1]}&\multicolumn{1}{c}{TI\footnotemark[2]}&\multicolumn{2}{c}{MM-PBSA\footnotemark[2]}&\multicolumn{2}{c}{This
    work}\\
  \multicolumn{1}{c}{Ligand}&$\Delta G_{\rm{exp}}$&$\Delta\Delta
  G_{\rm{exp}}$&$\Delta\Delta G_{\rm{bind}}$&$\Delta G_{\rm{(MM+solv)}}$&$\Delta\Delta G_{\rm{(MM+solv)}}$&$\Delta
  G_{\rm{calc}}$&$\Delta\Delta
  G_{\rm{calc}}$\\ \hline
  Theophylline&-8.85&0&0&-18.51&0&-16.1&0\\
  3-Methylxanthine&-7.77&1.08&1.36&-15.27&3.25&-14.7&1.4\\
  Xanthine&-6.91&1.94&1.64&-12.97&5.54&-14.6&1.6\\
  1-Methylxanthine&-6.88&1.97&1.82&-14.24&4.27&-14.2&1.9\\
  Hypoxanthine&-5.87&2.98&---&---&---&-11.6&4.5\\
  Caffeine&-3.35&5.50&6.63&-12.67&5.84&-11.0&5.1\\
\end{tabular}
\end{ruledtabular}
\footnotetext[1]{These values are obtained using the formula -$RT$ln~$K_{\rm d}$, where $K_{\rm d}$ is the individual competitor dissociation constant.~\cite{Jenison1994}. $T$=298~K.}
\footnotetext[2]{These values are from Ref.~\onlinecite{Gouda2003}.}
\end{table*}

\newpage
%%%%%%%%%%
% Fig. 1 %
%%%%%%%%%%

\newpage
%%%%%%%%%%
% Fig. 2 %
%%%%%%%%%%

\newpage
%%%%%%%%%%
% Fig. 3 %
%%%%%%%%%%

\newpage
%%%%%%%%%%
% Fig. 4 %
%%%%%%%%%%

\newpage
%%%%%%%%%%
% Fig. 5 %
%%%%%%%%%%
\begin{figure}
\includegraphics[width=14cm]{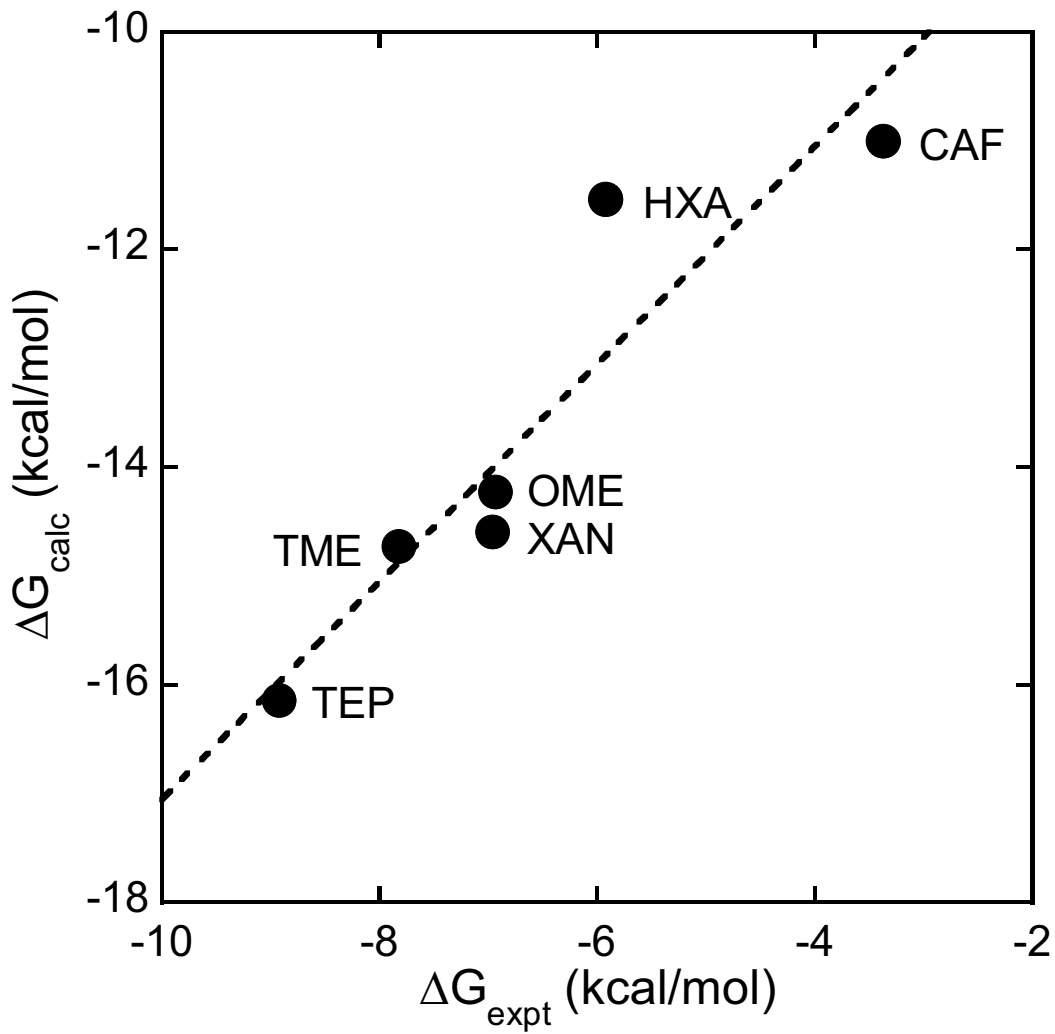}
\caption{\label{fig:DG_plot_6molecules} Computed binding free energies
  $\Delta G_{\rm calc}$ vs. experimentally measured binding free
  energies $\Delta G_{\rm expt}$ for six ligands. The line of slope= 1
  is also drawn as a guide. The abbreviations used here are TEP as
  theophylline, TME as 3-methylxanthine, XAN as xanthine, OME as
  1-methylxanthine, HXA as hypoxanthine and CAF as caffeine.
}
\end{figure}

\newpage
%%%%%%%%%%
% Fig. 6 %
%%%%%%%%%%

\end{document}